\documentclass[aps,prd,reprint,nofootinbib,superscriptaddress]{revtex4-2}
\usepackage{comment}
\usepackage{xcolor}
\usepackage{graphicx}
\usepackage{amsmath}
\usepackage{appendix}
\usepackage{footmisc}
\usepackage[english]{babel}
\usepackage[utf8]{inputenc}

\newcommand\keff{$\kappa_{\textrm{eff}}$}

\newcommand\veff{v$_{\textrm{eff}}$}

\newcommand{\sam}[1]{\textcolor{magenta}}

\newcommand\knaught{$\kappa_{\textrm{0}}$}
\begin{document}

\title{Time-Dependent Cosmic Ray Halos from Bursty Star Formation and Active Galactic Nuclei: Semi-Analytic Formalism and Galaxy Formation Implications}

\author{Sam B. Ponnada}
\affiliation{Department of Physics and Astronomy, Chalmers University of Technology, Gothenburg, Sweden}
\affiliation{TAPIR, California Institute of Technology, Mailcode 350-17, 
Pasadena, CA 91125, USA}

\email{sam.ponnada@chalmers.se}

\

\begin{abstract}
   Cosmic ray (CR) feedback in galaxy evolution has seen a theoretical resurgence in the past decade, but significant uncertainties remain in CR transport through the interstellar and circum-galactic media (ISM and CGM). While several works indicate CR effects may be notable in both star-forming and quenched massive galaxies, modeling the vast CR transport parameter space currently allowed by observations is computationally restrictive to survey. Analytic treatments of CR feedback have provided useful insights to potential ramifications in different regimes, but have relied on time-steady assumptions which may not well characterize CR effects at different cosmic epochs and galaxy mass scales. We present semi-analytic approximations and numerical solutions describing the time-dependent evolution of CR pressure in the CGM under simplified assumptions, which allow for quick evaluation of the vast allowable CR transport parameter space. We demonstrate that time-dependent injection from bursty star formation and/or episodic black hole accretion can substantially alter CR pressure profiles, particularly in the outer halos of massive galaxies ($\gtrsim R_{vir}$). Finally, we benchmark the approximate solutions from our semi-analytic formalism against a cosmic ray-magnetohydrodynamic (CR-MHD) cosmological zoom-in galaxy simulation directly modeling the CR scattering rate and emergent transport in full generality, highlighting the validity of our approach. We conclude by motivating careful consideration of time-dependent ``softening" effects in sub-grid routines for CR feedback, particularly for use in large cosmological volumes.
\end{abstract}

\maketitle

\section{Introduction} \label{sec:intro}
In recent years, it has become clear that in order to advance our understanding of feedback in galaxy formation, the details of feedback ``microphysics" must be modeled directly. Notably, there has been a resurgence of focus on one such ``microphysical'' source of feedback: cosmic rays (CRs) \citep[see][for a recent review]{ruszkowski_cosmic_2023}. Much work on CR feedback of late has focused on injection from supernovae (SNe) and resultant effects in galaxies at or below the break of the galaxy stellar mass function \citep{butsky_role_2018,Hopkins2020,quataert_physics_2022,pfrommer_simulating_2022,thomas_cosmic-ray-driven_2023,modak_cosmic-ray_2023,rodriguez_montero_impact_2024}. 

These works in large part have shown that for empirically-motivated CR transport parameters, CRs (particularly the $\sim$ GeV protons which dominate the bulk CR pressure) could have significant effects on galaxy growth and baryon cycling, influencing the bulk kinematics and phase structure of outflows and halo gas \citep{Ji2020,butsky_impact_2020,buck_effects_2020,butsky_impact_2022,ponnada_magnetic_2022,ponnada_synchrotron_2024}.

Above the break in the galaxy mass function, active galactic nuclei (AGN) are crucial to regulating galaxy growth and evolution, particularly at high dark matter halo masses (M$_{\textrm{halo}}$ $\gtrsim$ 10$^{13} M_\odot$; \citep{harrison_impact_2017}). Despite long-standing knowledge of AGN feedback's role at these mass scales \citep{croton_many_2006}, \textit{how} massive black holes' energy injection couples to the surrounding interstellar and circum-galactic medium (ISM and CGM) is largely unknown.

Many state-of-the-art simulations and semi-analytic models utilize variations on thermal and kinetic energy injection into AGN surroundings in order to effectively ``quench'' star formation in massive galaxies and reproduce observed galaxy properties \citep{schaye_eagle_2015,pillepich_simulating_2018}. While these approaches reify the importance of AGN, the physical nuances of feedback models remain a significant open question to be confronted with multi-wavelength observational constraints. 

Indeed, radio emission arising from CR electrons tracing AGN activity has long been studied as an important clue towards AGN feedback and its coupling to galactic environments \citep[see][for relevant reviews]{heckman_coevolution_2014,hardcastle_radio_2020}. Idealized simulations of massive galaxies \citep{su_which_2021,su_unraveling_2023,su_modeling_2025} have begun to explore the vital role CRs from AGN may play in the cessation of star formation and maintenance of quenching, in conjunction with other known feedback mechanisms.

Self-consistent, cosmological simulations of galaxy formation have also found that injecting a small, fixed fraction of the AGN accretion energy into CRs (marginalizing over the details of the injection physics on the much smaller accretion disk scales), produce reasonably quenched massive galaxies \citep{wellons_exploring_2023}, without obviously violating known observational constraints \citep{byrne_effects_2024,ponnada_hooks_2025}.

Although CR transport parameters, which depend on plasma microphysics on $\sim$AU scales \citep{zweibel_microphysics_2013}, are still unknown, a novel framework is emerging where for plausible transport prescriptions, CRs \textit{could be important for maintaining quenching} on longer timescales. Recently, \citep{quataert_cosmic_2025} demonstrated  using order-of-magnitude analytic arguments for plausible ``effective" CR diffusion/streaming speeds and fractional injection of AGN accretion energy into CRs, CRs may drive outflows on larger scales (beyond $\gtrsim$ R$_{vir}$) from group-mass halos (M$_{\textrm{halo}} \sim$ 10$^{13}$ M$_{\odot}$).

Analytic work modeling CR feedback from galaxies have typically focused on spherically-symmetric, steady-state wind solutions  \citep{ipavich_galactic_1975,quataert_physics_2022_diffusion,quataert_physics_2022_streaming,modak_cosmic-ray_2023,butsky_constraining_2023,hopkins_cosmic_2025} for spatially and temporally constant transport parameters and time-steady CR injection and pressure evolution, which have also been implemented as sub-grid models \citep{hopkins_simple_2023}. While these approaches are physically intuitive and are valid first approximations, star formation across galaxy mass scales can be highly time-variable \citep{muratov_gusty_2015,sparre_starbursts_2017}, and AGN accretion is notoriously episodic \citep{ulrich_variability_1997}. So, while steady-state approximations may well characterize the Milky Way and other low-$z$ spiral galaxies which have been continuously forming stars and thus steadily injecting CRs into their halos for the past several Gyr, these approximations may not hold for galaxies with burstier star formation histories (SFHs), highly episodic AGN injection, and/or galactic halos at high $z$ due to non-negligible source evolution and long travel times. 

Ultimately, the macroscopic physical quantities of interest for CRs vis-a-vis galaxy formation are the CR pressure, P$_{\textrm{CR}}$, and its gradient. Numerical experiments and galaxy-scale simulations which explicitly evolve CRs, however, remain very computationally expensive. Much like the radiative transfer problem, modeling physics with signal speeds $\sim\mathcal{O}(c)$ in systems where characteristic speeds are several dex slower on average introduces additional overhead on simulations which already suffer from having to evolve large spatial and temporal dynamic ranges for end-to-end predictions, even with ``reduced speed-of-light" methods. To ameliorate this issue, further robust analytic treatments and  numerical routines are required to advance our physical intuition and expand the prediction space for CR physics in galaxy formation.

Towards this end, we analytically and numerically explore CR feedback from star-forming and massive galaxies. Specifically, we relax the common assumption in the literature of time-steady CR injection and pressure evolution ($\S$ \ref{sec:sec2} \& $\S$\ref{sec:sec3}), to explore semi-analytic and numerical solutions to CR transport in halos for diffusion and streaming/advection-dominated regimes. We provide a generalizable framework for the evolution of CR pressure profiles in galaxy halos, independent of the choice of transport parameters or injection cadence. To benchmark our simple, yet surprisingly viable semi-analytic approach, we compare against ``true'' numerical solutions, as well as a fully cosmological, self-consistent, cosmic-ray-magnetohydrodynamic (CR-MHD) zoom-in simulation of a massive galaxy ($\S$\ref{sec:validation}). ($\S$\ref{sec:discussion_conclusions}) summarizes our results and motivates improved sub-grid modeling of CR physics in large-volume cosmological simulations including AGN feedback, for which we will present open-source numerical tools in future work.

\section{Analytic Expectations for Constant Cosmic Ray Transport Parameters}\label{sec:sec2}

The transport of CRs through the ISM and into the CGM of massive halos is fundamentally a multi-scale problem connecting the details of ``micro-physics" on gyro-resonant scales r$_{\textrm{gyro}}$ $\lesssim$ AU in typical galactic magnetic field strengths $\sim$ \textbf{$|B|/\mu G$} to ``macro-scopic'' ($\sim$ kpc) scales.

However, the micro-physics which determine the pitch-angle scattering rate of CRs $\nu_{\textrm{CR}}$, giving rise to \textit{effective} `streaming' or `diffusive' behavior in varied limits of $\nu_{\textrm{CR}}$, remain unknown. Thus, we may parameterize our ignorance of micro-physics for the relevant transport through the emergent, large-scale diffusion and advection/``streaming'' coefficients $\kappa_{\textrm{0}}$ and v$_{\textrm{st,eff}}$ (v$_{\textrm{eff}}$ hereafter, defined further below), as done in previous works \citep{Hopkins2020,butsky_constraining_2023,hopkins_simple_2023,hopkins_cosmic_2025}. So, we treat CR transport here in the relativistic fluid limit with the CR pressure given by P$_{\textrm{CR}} = (\gamma_{\textrm{CR,\, ad}}-1)\, e_{\textrm{CR}}$, where e$_{\textrm{CR}}$ is the CR energy density, and $\gamma_{\textrm{CR,\, ad}} = 4/3$ is the CR adiabatic index. Here, e$_{\textrm{CR}}$ implicitly represents the spectrum-integrated CR energy density, dominated by the $\sim$ GeV CR protons  

We then assume that a fixed fraction of AGN accretion energy $\epsilon_{\textrm{CR, BH}}$, or of SNe shocks per unit star formation ($\sim$ 0.1 * 10$^{51}$ erg/100 M$_{\odot}$), $\epsilon_{\textrm{SF}}$, is converted into CRs. The rate of CR energy injection is given by

\begin{equation}\label{eq:ECR_accretion}
     \dot{E}_{\mathrm{CR}}(t) = \epsilon_{\mathrm{CR,BH/SF}} \,\dot{M}_{\mathrm{BH/SF}}(t)\,c^2 
 \end{equation}
 
 The injection occurs as a ``pulse'' with infinitesimal width represented as a spatial and temporal $\delta$-function initially. Then, the CR pressure source term $S$ becomes
 \begin{equation}\label{eq:point_source_Pdot}
     S = \dot{P}_{\mathrm{CR}}(t) = (\gamma_{\textrm{CR,\, ad}}-1) \,  \dot{E}_{\mathrm{CR,BH/SF}}(t)\, c^{2} \delta(\vec{r}, t)
 \end{equation}
 
 where $\vec{r}$ is the vector position from the central BH, and $t$ represents the time the pulse was injected. If such a pulse of CRs  experience post-injection losses due to hadronic collisions, ionization, Coulomb interactions, and so on, the ``calorimetric fraction" can be written as f$_{\textrm{cal}}$, where pure calorimetry means f$_{\textrm{cal}}$ = 1, i.e., no CRs escape from the source region into the CGM. 
 
 We further impose spherical symmetry and solve for the Green's function solutions to the 3D diffusion-advection equation of the following form, assuming magnetic fields are isotropically tangled on large spatial scales in the CGM \citep[which simulations indicate is a reasonable approximation][]{Ji2020,ponnada_magnetic_2022} \footnote{Note, in practice the CR transport equations are not formally a diffusion-advection equation, but implicit in our assumptions here is treating the `streaming' and `diffusive' limits of micro-physical scattering as \textit{effective} fluid transport speeds on scales $>>$ the CR scattering mean free path, which then gives rise to the resulting CR pressure. Exact separation of CR transport into `streaming-like' vs. `diffusive' behavior is only possible when these coefficients are constants, and otherwise become strictly degenerate once their true arbitrary scalings with various locally varying plasma properties are considered (see \citep{hopkins_testing_2021} Appendix B, and \citep{hopkins_simple_2023} for more detailed discussions). \label{footnote1}}

\begin{equation}\label{eqn:general_PDE}
    \begin{split}
    \tiny
            \frac{\partial P_{\mathrm{CR}}(\vec{r}, t)}{\partial t}
        = \\ & \tiny \nabla \cdot \Big[\kappa_{\textrm{0}}(r) \nabla P_{\rm CR} - v_{\textrm{eff}}(r) P_{\rm CR}\Bigr] \\ & \tiny + (1-f_{\textrm{cal}})\,S(t)\, \\ & - \Lambda_{\textrm{st}/ad}(r,t)
    \end{split}
\end{equation}

where v$_{\textrm{eff}}$ is the magnitude of the effective advection/convection + streaming velocity $\textbf{v}_{\textrm{eff}}= \textbf{u}_{\textrm{gas}}$ + v$_{\textrm{A}} \hat{\nabla}P_{\textrm{CR}}$. $\Lambda_{\textrm{st}/ad}$ is the streaming + adiabatic loss term ($P_{\textrm{CR}}(\nabla \cdot \textbf{v}_{\textrm{eff}})$) with v$_{\textrm{A}}$ the local Alfv\'en speed and $\kappa_{\textrm{eff}}$ the  local isotropic diffusive transport coefficient emerging from microphysical scattering on gyro-resonant scales. 

The terms in $\Lambda_{\textrm{st}/ad}$ can be expanded to $\Lambda_{\textrm{st}/ad} = $ $[\frac{P_{\textrm{CR}}}{3} \nabla \cdot \textbf{v}_{\textrm{eff}} + \frac{2 v_{\textrm{eff}} P_{\textrm{CR}}}{3r}]$. Note here we have taken the isotropic-equivalent diffusion coefficient ($\kappa_{0} \sim \kappa_\|/3)$ and solely the radial components in $\textbf{v}_{\textrm{eff}}$ for our spherically symmetric assumptions here. This means we implicitly assume the transport coefficients averaged over large spatial scales sampled by CR travel paths along tangled field lines gives rise to approximately isotropic, radial ``diffusive'' and ``streaming/advective-like" motion, though the equations presented here remain agnostic to local non-radial components in $\hat{\nabla}P_{\textrm{CR}}$ on unresolved scales as in \cite{hopkins_simple_2023}. 

Furthermore, note that even if CRs ``stream" super-Alfv\'enically, the streaming losses must still be limited to the local Alfv\'en speed. Since we do not present calculations coupled to hydrodynamics in this work, we do not consider adiabatic and streaming losses with the exception of including the streaming term in the numerical solutions of Eq. \ref{eqn:general_PDE} in Fig. \ref{fig:numerical_validation}, taking v$_{\textrm{A}}$ estimated from the CR-MHD simulation we compare against. There, we show it makes a minimal difference, as the predicted v$_{\textrm{A}}$ at large CGM radii from full MHD simulations \cite{ponnada_magnetic_2022,ramesh_circumgalactic_2023} is small compared to the bulk CR transport speeds invoked.

On the other hand, if we were to include adiabatic losses based on a chosen v$_{\textrm{eff}}$ (or assumed u$_{\rm gas}$) here, it would represent an upper-limit to the adiabatic losses since, by construction, we do not capture the adiabatic back-reaction by gas onto the CR fluid. Moreover, the non-linear nature of these losses would make finding simple closed-form expressions to the time-dependent problem intractable. We will further discuss losses in \S \ref{sec:discussion_conclusions}.


In essence, our parameterization separates the scalar transport behavior into ``diffusion-like" and ``streaming + advection-like" regimes (see footnote \footref{footnote1}). Below, we delineate how the normalization and shape of the CR pressure profile is sensitive to the injection history of CRs and transport governed by \knaught\, and/or \veff.

\subsection{Case 1: Constant $\kappa_{0}$, No Advection or Streaming}\label{sec:case1}

If $\kappa_{0}$ is a constant value due to an effectively constant $\nu_{\textrm{CR}}$) throughout the ISM and CGM as is commonly assumed for simplicity in most state-of-the-art, CR-MHD galaxy simulations (and when averaged over large spatial scales as we are assuming here) \citep{butsky_role_2018,chan_cosmic_2019,Hopkins2020,farcy_radiation-magnetohydrodynamics_2022,hopkins_first_2022,rodriguez_montero_impact_2024,ponnada_hooks_2025}, and \veff\, $\rightarrow$ 0, Equation \ref{eqn:general_PDE} reduces to a diffusion equation with a Green's function solution with a Gaussian form:

\begin{equation}\label{eqn:PdotCR_no_adv_const}
    \begin{split}
    P_{\mathrm{CR}}(\vec{r}, t) =\\
    & \int^{t}_{0} \frac{\dot{E}_{\textrm{CR}}(t-t')}{3\bigl(4\pi \kappa_{\textrm{0}}\,t'\bigr)^{3/2}}
    \exp\!\Biggl(
       -\,\frac{r^{2}}{4\,\kappa_{\textrm{0}}\,t'}
    \Biggr) dt'
    \end{split}
\end{equation}

Here, r is the scalar galactocentric distance in spherical coordinates, and t' is the time since injection of a given pulse, or the look-back time to the injection at time t$_{\textrm{i}}$ ($t'$ $\equiv$ t - t$_{\textrm{i}}$).

Now, if $\dot{M}_{\textrm{BH/SF}}$ is dominated by a single injection event, this Gaussian expression is the exact closed-form solution. This is shown in \citet{quataert_cosmic_2025}, where the authors consider constant-\knaught\, models for CR energy injection from AGN, with a small, fixed fraction of accretion energy converted to CRs. There, the authors find that for isotropically-averaged \knaught\, $\sim$ 10$^{ 30} \textrm{\, cm}^2\, \textrm{s}^{-1}$, P$_{\textrm{CR}}$ can be approximately in equipartition with, or dominant to the thermal pressure P$_{\textrm{th}}$ around the virial radius of massive galaxies ($M_{\textrm{halo}} \gtrsim$ 10$^{13} M_{\odot}$) with similar assumptions as above, comparing to empirical thermal pressure profiles for galaxy groups and clusters from X-ray observations \citep{arnaud_universal_2010}. 

Otherwise, when $\dot{M}_{\textrm{BH/SF}}$ are more complex, Equation \ref{eqn:PdotCR_no_adv_const} represents the convolution of the Green's function solution with the time dependent source term. In global flux steady-state, for the same spherically symmetric assumptions here, a simple closed form solution of $P_{\textrm{CR}} = \frac{\dot{E}_{\textrm{CR}}}{12\pi\,\kappa_{\textrm{0}}\,r} $ can be found, as has been widely adopted in the literature \citep{Hopkins2020,hopkins_testing_2021,hopkins_effects_2021,quataert_physics_2022_diffusion,butsky_constraining_2023}. This solution is valid out to the effective diffusion radius $\sim \sqrt{\kappa_{\textrm{0}} \tau}$, with a significant exponential tail extending out to $\sim \sqrt{4\kappa_{\textrm{0}} \tau}$, where $\tau$ represents the time over which CRs have been injected.

\subsection{Case 2: \veff-dominated Transport}\label{sec:case2}

In the limit of \knaught\, $\rightarrow 0$, CR transport is dominated by advective outflow and/or by streaming, and Equation \ref{eqn:general_PDE} reduces to the advection equation in spherical coordinates. In global flux steady-state, this has a well-known solution of P$_{\textrm{CR}} = \dot{\textrm E}_{\textrm{CR}}/(12\pi v_{\textrm{eff}} r^2)$ \citep{quataert_physics_2022_streaming,quataert_physics_2022_diffusion,hopkins_cosmic_2025} out to some finite travel distance $\sim$ \veff t.

We now explore how time-dependence of injection affects this solution. First, consider a strongly peaked injection history around z $\sim 2-3$ as expected for peak star formation or z $\sim 1-2$ for peak massive black hole accretion. This injection then falls off with absolute time as $\sim t^{- \xi}$, where $\xi$ is a power-law fit to the falling (but bursty) low-redshift injection history. 

Now, if $\dot{\textrm{E}}_{\textrm{CR}} \sim t^{- \xi}$, and the injected pulse propagates primarily at v$_{\textrm{eff}}$, then $\dot{\textrm{E}}_{\textrm{CR}}$ increases with $t'$, the look-back time since injection along the characteristic $t'$ = r/v$_{\rm eff}$, as larger look-back times are closer to the peak of injection. Thus, the expected P$_{\textrm{CR}}$ profile in this limit should be shallower than the r$^{-2}$ behavior of the steady-state wind case, instead being $\propto r^{-2+\xi}$ to leading order. This holds when $\dot{E}$ varies significantly on a timescale $\lesssim$ the transport time r/v$_{\textrm{eff}}$. 

Formally, a time-dependent solution to this advection-like problem can be worked out via the method of characteristics for a constant \veff, where r(t) = r$_{0}$ + \veff $\ast$ t. Explicitly including the time dependence within the boundary condition and assuming there is no shell mixing leads to 

\begin{equation}\label{eq:Adv_only_td}
    P_{\textrm{CR}}(r,t) =  \frac{ \dot{\textrm{E}}_{\textrm{CR}}(t-t')}{(12\pi v_{\textrm{eff}} r^2) } H(t-t')
\end{equation}

where for the advection-only case $t' \equiv \frac{r}{v_{\textrm{eff}}}$ strictly and H is the Heaviside step function. Essentially, the pressure profile at a given radius will depend on the CR injection at $t-t'$, and vanishes beyond the finite travel-time distance.

\subsection{Case 3: Constant \knaught, Constant \veff}\label{sec:case3}

When the transport of CRs is not solely in the diffusive limit, we must account for the effective streaming and/or advection speed. In this scenario, for uniform \veff\,  an exact solution to Eq. \ref{eqn:general_PDE} in Cartesian coordinates can be found,  

\begin{equation*}\label{eqn:PdotCR_const_diff_adv}
    \begin{split}
    P_{\mathrm{CR}}(\vec{r}, t) = \\
    & \int^{t}_{0} \frac{\dot{E}_{\textrm{CR}}(t-t')}{3\bigl(4\pi \kappa_{\textrm{0}}\,t'\bigr)^{3/2}}
    \exp\!\Biggl(
       -\,\frac{\|\mathbf{r}-\mathbf{v_{eff}}t'\|^{2}}{4\,\kappa_{\textrm{0}}\,t'} 
    \Biggr)dt'
    \end{split}
\end{equation*}

wherein \textbf{r} and \textbf{v$_{\textrm{eff}}$} represent the vector position and effective advection and/or streaming speed. However, proper treatment of Eq. \ref{eqn:general_PDE} in spherical coordinates forgoes such an exact and simple solution via Galilean transformation of a Gaussian. This owes to the extra curvature terms of form $\frac{2}{r} \nabla P$ introduced by the divergence in spherical coordinates. The far-field regime (large r) is precisely where these curvature terms will be less significant for the shifted Gaussian solution, though it will strictly over-estimate the true total pressure at inner to intermediate radii and subsequently not conserve total energy as the exact solution should.

We can approximate the true form for constant \knaught\, and \veff, by ignoring the curvature terms, finding

\begin{equation}\label{eqn:PdotCR_const_diff_adv_approx}
    \begin{split}
    P_{\mathrm{CR}}(\vec{r}, t) \approx \\
    & \int^{t}_0 \frac{\dot{E}_{\textrm{CR}}(t-t')}{3\bigl(4\pi \kappa_{\textrm{0}}\,t'\bigr)^{3/2}}
    \,\exp\!\Biggl(
       -\,\frac{(r- v_{\textrm{eff}}t')^{2}}{4\,\kappa_{\textrm{0}}\,t'}
    \Biggr) dt'
    \end{split}
\end{equation}

While strictly approximate, this functional form provides intuition for the true diffusion and streaming/advection solution, wherein as shells expand and move outwards from the central source, they will mix owing to diffusion, with the shells associated with a single injection event being ``smeared" out in radius as they propagate outwards.

However, it is possible to partially rectify the over-estimation to leading order by noting Eq. \ref{eqn:PdotCR_const_diff_adv_approx} represents the convolution of a time-dependent source function with a spatial Green's function kernel. 

Thus, we reformulate Eq. \ref{eqn:PdotCR_const_diff_adv_approx} as
\begin{equation}\label{eqn:SAM_econs}
    P_\textrm{CR}(\vec r, t) = \int_0^t \dot E_\textrm{CR}(t - t')\, A(t')\, \tilde g_0(r, t')\, dt',
\end{equation}

where
$$\tilde g_0(r, t') = \frac{\exp\!\left[-(r - v_{\rm eff} t')^2/(4\kappa_\textrm{0} t')\right]}{3(4\pi\kappa_\textrm{0} t')^{3/2}}$$
is the spatial Green's function kernel and
$$A(t') \equiv \left[\int_0^{R_\textrm{max}} 12\pi r^2\, \tilde g_0(r, t')\, dr\right]^{-1}$$
is the time-dependent normalization function. The form of $A(t')$ ensures the approximate solution conserves the total energy injected at any fixed time interval.

We enforce this normalization condition in the proceeding results, and we discuss and show later how this simple, approximate solution can be used as a semi-analytic model which matches the true numerical solution to Eq. \ref{eqn:general_PDE} to a surprisingly high degree of accuracy.

Note that in steady-state for the pressure equation, similarly to as in Case 1, a simple closed form solution of $P_{\textrm{CR}} = \frac{\dot{E}_{\textrm{CR}}}{12\pi\,\kappa_{\textrm{eff}}\,r} $ can be found, where $\kappa_{\textrm{eff}}$ subsumes the behaviors of $\kappa_{\mathrm 0}$ and v$_{\textrm{eff}}$ by defining $\kappa_{\textrm{eff}}$ for radii r $\geq$ r$_{\textrm{st}} = \kappa_{\mathrm 0}/v_{\textrm{eff}}$, \keff\, = $\kappa_{\mathrm 0}  + v_{\textrm{eff}} * r$ as an ``effective diffusion coefficient" (or conversely, one can write this an effective bulk streaming/advection speed). We will investigate the limitations of such simplifications in capturing time-dependent effects in the proceeding sections.

\subsection{Contrasting Constant CR Transport Parameter Behavior with Injection History}

The evolution of the halo CR pressure from an arbitrary injection history can be modeled using the approaches in the previous subsections. In this section, we explore the resulting variation in spatial distribution of CR energy for different injection histories.

\begin{figure}
    \centering
    \includegraphics[width=1.0\linewidth]{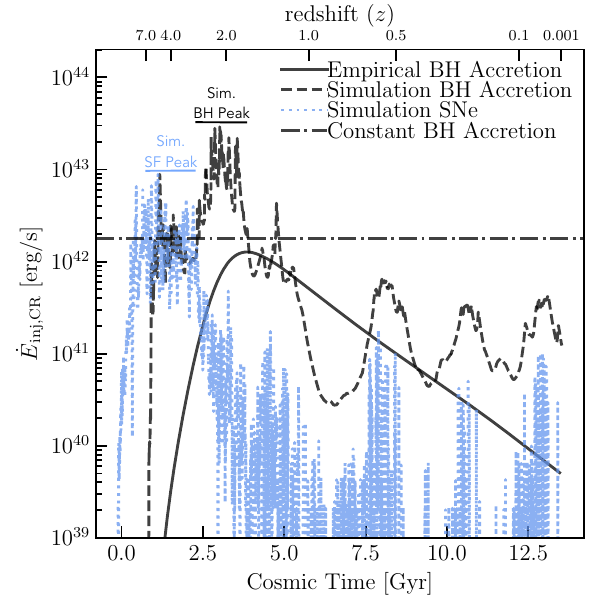}
    \caption{\textit{Reference CR energy injection histories used for our model comparisons in this study.} The simulation injection histories from BH and supernova contributions (black dashed and blue dotted) are taken from a fully dynamical CR-MHD simulation of a massive halo from the FIRE-3 simulation suite \citep{hopkins_fire-3_2023,byrne_effects_2024, ponnada_hooks_2025}, with the peak injection from star formation and black hole accretion dermarcated. The constant BH accretion model takes the total energy integrated over time for the simulation model history and averages over a Hubble time. For each BH accretion cases, we assume $\epsilon_{\mathrm{CR,BH}} = 3 \times 10^{-4}$.The empirically motivated injection history (solid line) follows the average black hole accretion rate for M$_{\textrm{halo}}^{z=0} = 10^{13} M_\odot$ halos from the empirical model Trinity \citep[Fig. 16 of][]{zhang_trinity_2023}.}
    \label{fig:injection_histories}
\end{figure}

We first show the contrasting model injection histories we explore in Figure \ref{fig:injection_histories}. There are four model injection histories considered: (1) a single large delta function injection from BH accretion peaked at z$\sim$3 akin to \citet{quataert_cosmic_2025}, (2) a time-dependent injection following the Trinity semi-empirical model for the cosmic-averaged BH accretion histories of M$^{z=0}_{\rm halo} = 10^{13} M_\odot$ galaxies \citep[Fig. 16 of][]{zhang_trinity_2023}, (3) a bursty, time-dependent injection history from SNe and BH accretion taken from a cosmological, CR-MHD zoom-in simulation from the Feedback in Realistic Environments (FIRE-3) suite\footnote{\protect\url{https://fire.northwestern.edu/}} of a massive galaxy (M$^{z=0}_{\rm halo} \approx 10^{13} M_\odot$ )\citep{hopkins_fire-3_2023,ponnada_hooks_2025,goyal_effects_of_CR_AGN_2026}, and (4) a constant, time-steady injection history which averages the total CR energy injected by the BH in the simulation over a Hubble time. The empirical injection history marginalizes over short time-scale variability that would be seen in individual halos, as it represents an ensemble average. We aim to contrast this smoothed scenario with the bursty, simulated injection history. 

\begin{figure*}
    \centering
    \includegraphics[width=1.0\linewidth]{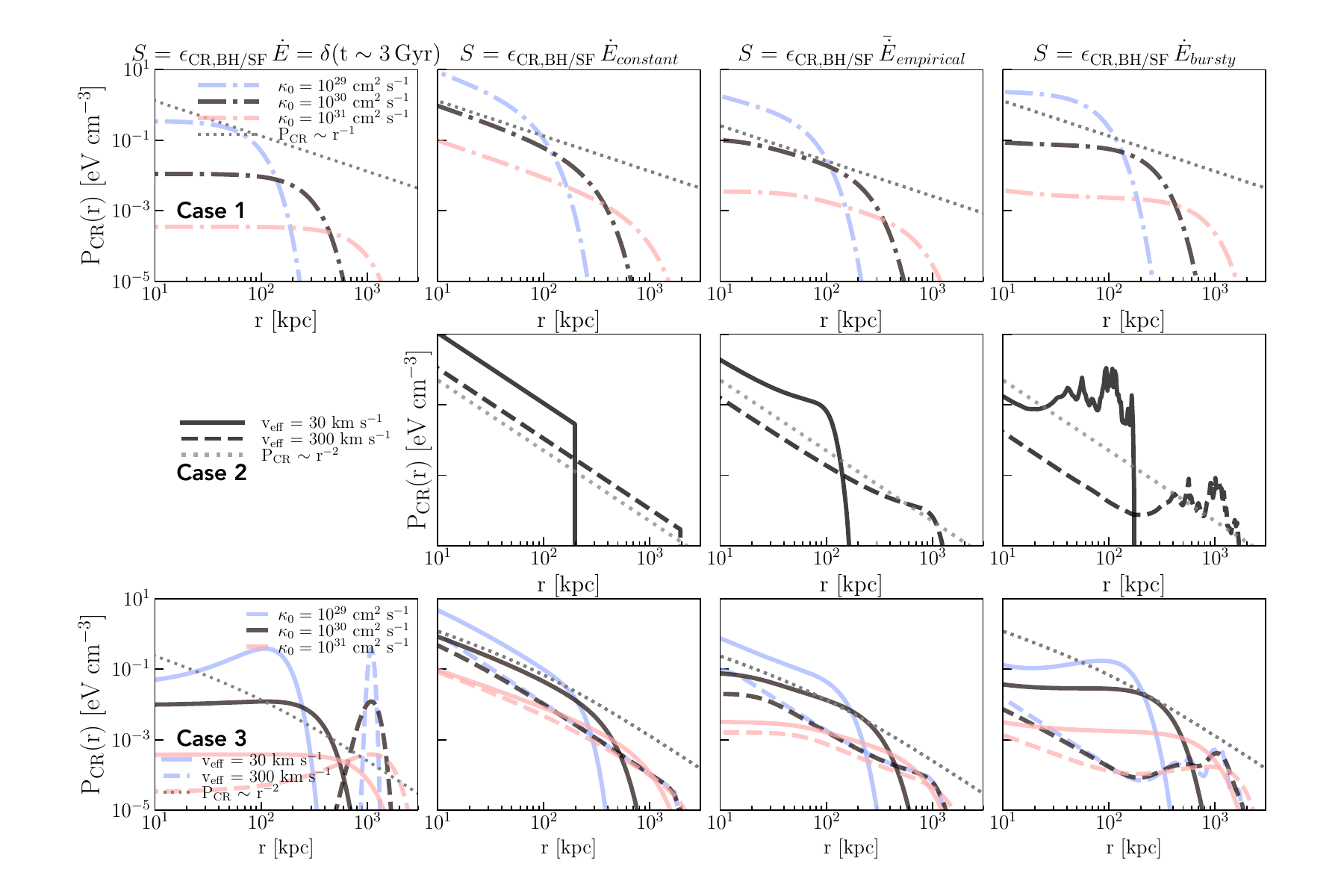}
    \caption{\textit{Exact diffusion-only (Case 1; Eq. \ref{eqn:PdotCR_no_adv_const}), advection/streaming-only (Case 2; Eq. \ref{eq:Adv_only_td}, and approximate, semi-analytic diffusion+streaming/advection (Case 3; Eq. \ref{eqn:SAM_econs}) solutions for P$_{\textrm{CR}}$ in a massive galaxy halo (M$^{z=0}_{\textrm{halo}} = $ 10$^{13} M_\odot$) at $z = 0.8$ with $\epsilon_{\mathrm{CR,BH}} = 3 \times 10^{-4}$, for varied injection histories (left to right).} We show solutions for a single, strongly peaked $\delta$-function injection at $z\sim3$ (\textbf{1st column}), a constant injection history over the Hubble time up to $z = 0.8$ (\textbf{2nd column}), a peaked but slowly decaying CR injection history based on the Trinity empirical model for the average black hole accretion rate \citep{zhang_trinity_2023} (\textbf{3rd column}), and a bursty, time-dependent CR injection history from a simulated cosmological zoom-in CR-MHD simulation (\textbf{4th column}). The colors in Cases 1 and 3 denote increasing \knaught\, (10$^{29}$, 10$^{30}$, 10$^{31}$) cm$^{2}$ s$^{-1}$ in blue, black, and pink, respectively, with different line-styles in Case 2 and 3 (solid, dashed) denoting varied \veff\, (30, 300) km s$^{-1}$. Example infinite time steady-state solutions are shown with dotted gray lines in all panels. Compared to a single $\delta$-injection, time-dependent injection can shift the spatial distribution of CR energy, increasing the pressure at inner radii for the same transport parameterization due to late-time injection, where compared to steady injection, both the smoothly falling and bursty time-dependent injection flatten profiles at large radii relative to steady-state expectations.}
    \label{fig:const_kappa_models}
\end{figure*}

In Figure \ref{fig:const_kappa_models}, we compare all injection histories shown in Figure \ref{fig:injection_histories} for each outlined Case in Section \ref{sec:sec2}. To simplify the qualitative comparison here, we consider only injection from black hole accretion with the same $\epsilon_{\mathrm{CR,BH}} = 3 \times 10^{-4}$ for each case. As an example, we show the behaviors of the Case 1-3 pressure profiles integrated over cosmic time up to z $\sim$ 0.8. We do not show the $\delta$-injection result for Case 2 as it is simply a traveling pulse. Also note for the Case 3 pressure profiles, we follow the normalization scheme of Eq. \ref{eqn:SAM_econs} for each toy-model solution to approximate the energy-conserved solution, and is shown here to demonstrate the qualitative behavior of a diffusion + streaming/advection solution. We will demonstrate the soundness of this approximation in  \S\ref{sec:sec3}.

When examining the second row of Figure \ref{fig:const_kappa_models}, two behaviors immediately become clear. First -- effective streaming/advection (for plausible speeds) moves the bulk of the CR energy outwards relative to the pure diffusion case. This is clear not only from the translation of the pressure fronts outwards, but from the flattening of the profiles at larger radii. These trends imply that for non-negligible streaming/advection speeds \textit{CR pressure may not only be important at the virial radius of massive galaxies \citep{quataert_cosmic_2025}, but even out to larger, cosmologically relevant scales} ($\geq$ R$_{\rm vir} \sim$ 300 kpc).

Secondly, \textit{time dependent injection can significantly alter CR pressure profiles}. This is evinced by the changes apparent for the same toy model transport parameters between the columns of Figure \ref{fig:const_kappa_models}. In particular, compared to a singular delta function source, time-dependent injection naturally elevates the pressure profiles in the inner halo ($r \lesssim 100$ kpc) due to late-time injection beyond z $\sim 3$, and also serves to flatten the profiles at large radii relative to steady-state scalings described in \S \ref{sec:sec2}. 

The isolated effect of time dependence is especially clear in the 2nd row of Figure \ref{fig:const_kappa_models}, where there is no diffusion to smooth out the pressure at large radii -- any flattening relative to the analytic expectation of a $\sim$ r$^{-2}$ profile owes solely to the time dependence of injection, as we discussed in Section \ref{sec:case2}. Such flattening of the profiles even in the empirical injection case, which by construction averages over short fluctuations in the injection rate, indicates that even the slow evolution of the \textit{average} black-hole accretion rate (or cosmic star formation rate) can substantially influence the distribution of CR pressure in low-$z$ halos. Of course, if the injection history features large variation with time, this will naturally appear in the pressure profiles as seen in the right-most panels of Fig. \ref{fig:const_kappa_models}. In the presence of diffusion, the individual injection `bursts' get smeared across radial shells, though large amplitude fluctuations (depending on the diffusivity) may not get fully smeared into the flattened bulk profile (see the structure of the lowest diffusivity line in Fig. \ref{fig:const_kappa_models}, Case 3 right-most panel). 

In the following sections, we compare these exact and approximate analytic estimations to exact numerical solutions of Eq. \ref{eqn:general_PDE} and validate our simple numerical model for the evolution of CR pressure against full CR-MHD galaxy simulations.

\section{Numerical solutions to time-dependent CR Pressure Evolution for large scale diffusion and advection/streaming}\label{sec:sec3}

In this section, we present exact numerical solutions to Eq. \ref{eqn:general_PDE} under non-zero diffusion and advection/streaming to compare against our approximate Case 3 solutions. We show our approximations robustly describe the evolution of time-dependent CR pressure to leading order in galaxy halos.

To solve Eq. \ref{eqn:general_PDE}, we utilize a finite volume approach on a logarithmic radial grid of N$_{\textrm{grid}}$ cell centers evenly spaced between log$_{10}$(r/kpc) = -2 to 3.5, with a zero Neumann and outflow boundary condition at the inner and outer boundaries respectively. In this parameterization, the first grid cell implicitly contains r = 0, and P$_{\textrm{CR}}$ is initialized to 0 in each grid cell. We model the delta function source term by injecting $(\gamma_{\textrm{CR}}-1)$$\dot{E}_{\textrm{CR}}[t]$ in the first grid cell, normalized by the first cell's volume. The diffusive and effective streaming/advective fluxes are computed between  cell faces (assuming spherical symmetry) in a strictly energy conserving manner, with the evolved property being P$_{\textrm{CR}, i}$ = $(\gamma_{\textrm{CR}}-1)$$E_{\textrm{CR},\, i}/V_i$, the volumetrically-averaged CR pressure within each cell. 

For time integration, we use an implicit Runge-Kutta method of order 5 from the Radau IIA family \citep{hairer_stability_1996,hairer_stiff_1999}, implemented in \texttt{scipy} \citep{2020SciPy-NMeth}. This implicit method is suited for stiff problems like fast diffusion, and allows for larger time-stepping on high-resolution grids compared to explicit Runge-Kutte methods of the same order.

\begin{figure*}
    \centering
    \includegraphics[width=1.0\linewidth]{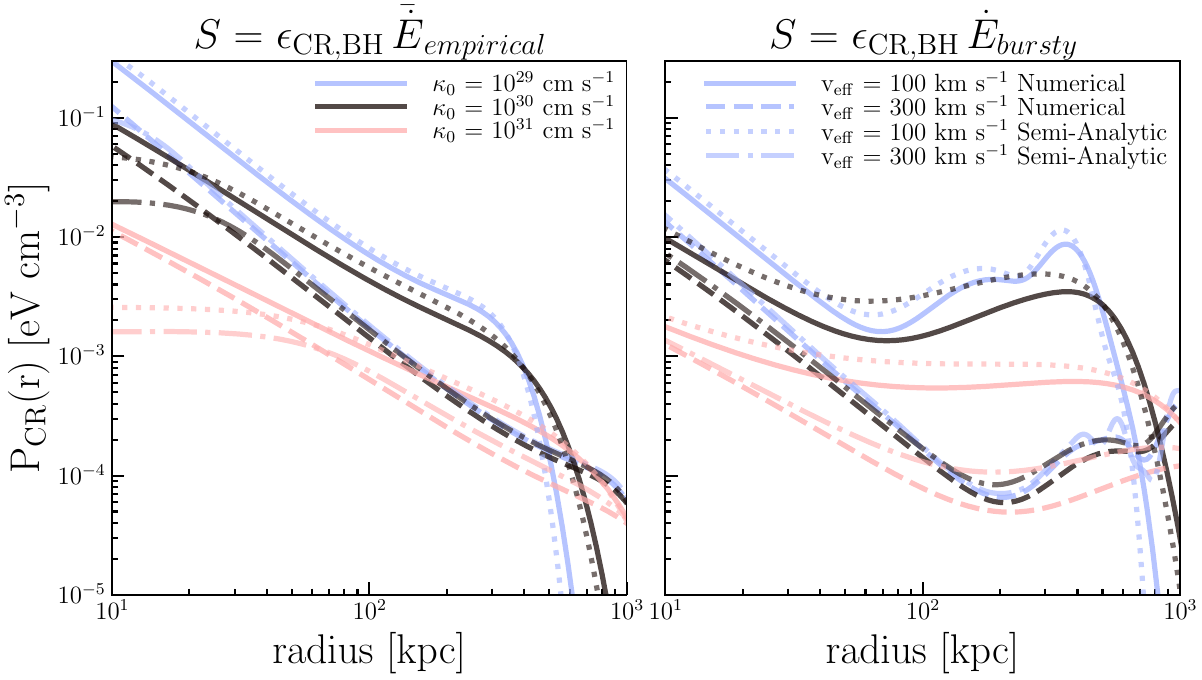}
    \caption{\textit{Exact numerical and approximate semi-analytic solutions for P$_{\textrm{CR}}$ (Eq. \ref{eqn:general_PDE}) in galactic halos at $z = 0.8$.} \textbf{Left:} Solutions for a peaked but slowly decaying CR injection history for a massive halo taken from an empirical model for the average black hole accretion rate \citep{zhang_trinity_2023}, with $\epsilon_{\mathrm{CR,BH}} = 3 \times 10^{-4}$. 
    \textbf{Right:} The same solutions for a bursty, time-dependent CR injection history for a simulated halo of the same mass with the same injection efficiency. In both panels, colors denote \knaught\, (10$^{29}$, 10$^{30}$, 10$^{31}$) cm$^{2}$ s$^{-1}$ in order of blue, black, pink, with solid and dashed lines denoting different \veff\, (100, 300) km s$^{-1}$, with the same shown for the approximate semi-analytic solutions in dotted and dot-dashed lines respectively . The true numerical solutions verify that time-dependent injection shifts the distribution of CR energy out to larger radii, which is further enhanced due to non-negligible effective streaming/advection relative to diffusion-only cases (c.f. Fig. \ref{fig:const_kappa_models}), and the approximate solutions show a high degree of validity, particularly at large radii, but with slight overestimation relative to the numerical solutions at intermediate radii and slight under-estimation at very inner radii, for the smoother injection case}.
    \label{fig:const_kappa_models_numerical}
\end{figure*}

In Figure \ref{fig:const_kappa_models_numerical}, we show the resulting numerical solutions to the Case 3 (diffusion + advection/streaming) CR transport for the same time-variable injection histories as in Figure \ref{fig:const_kappa_models}, evaluated out to z = 0.8. We limit the comparison here to the time-variable injection scenarios since the time-steady injection case has exact analytic solutions, discussed prior. Also, when we are working with the exact numerical solutions, we do not re-scale the corresponding profile in any manner -- the curves are manifestly energy conserving and any loss of energy simply represents the transport of flux out of the domain (f$_{\textrm{cal}}$ = 0 is assumed). We again assume $\epsilon_{\mathrm{CR,BH}} = 3 \times 10^{-4}$ for each case, and only include injection from BH accretion.

Like our approximate solutions to Case 3 in Figure \ref{fig:const_kappa_models} demonstrated, the true solution to the diffusion + advection/streaming equation in the presence of non-negligible time dependence and \veff\, results in more CR energy being shifted out to large radii, with the P$_{\textrm{CR}}$ profiles for bursty injection retaining some of the burst structure for large injection events and lower values of \knaught, but increasingly `losing memory' of the burst structures as diffusion becomes more important. Relative to the single $\delta$ injection approximation, CRs injected at later times boost the inner halo (r $\lesssim$ 100 kpc) profiles substantially. 

Comparing our semi-analytic formalism of Eq. \ref{eqn:SAM_econs} to the true numerical solutions shown in Fig. \ref{fig:const_kappa_models}, we find agreement to within tens of percent at large radii in both time-dependent cases, but with slight over-estimation of the "true" solution by Eq. \ref{eqn:SAM_econs} at intermediate radii and under-estimation at inner radii. The degree of this spatial discrepancy seems to be more pronounced for higher transport speeds (higher $\kappa_0$ and or \veff), which may indicate the approximation we made in Eq. \ref{eqn:SAM_econs} starting to break down in the limit of fast transport, even though the total energy injected remains conserved in both the exact numerical and approximate semi-analytic solutions. As we will show later in our comparison against a full, explicit CR-MHD simulation, these limitations still allow a surprisingly reasonable reproduction of a true simulated pressure profile for the proof of-concept we present here.

\begin{figure*}
    \centering
    \includegraphics[width=1.0\linewidth]{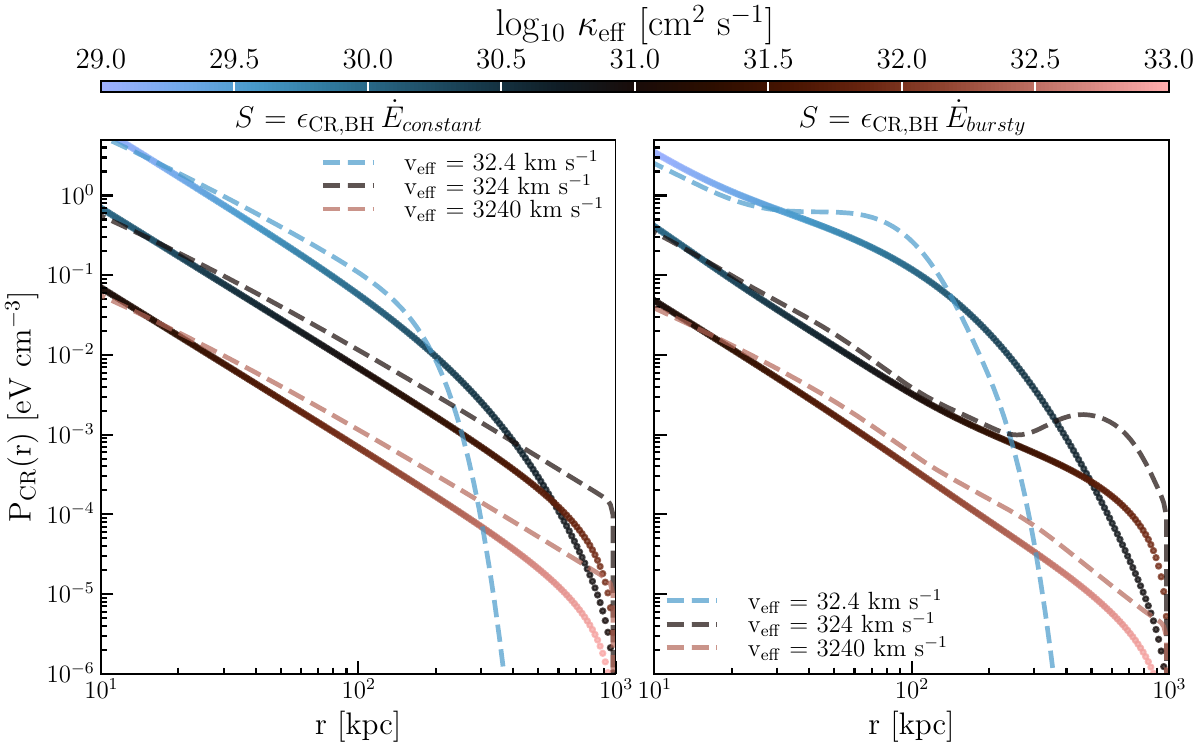}
    \caption{\textit{Numerical solutions for P$_{\textrm{CR}}$ (Eq. \ref{eqn:general_PDE}) in a massive galactic halo ($M_{\textrm{halo}}^{\textrm{z=0}}$ = 10$^{13} M_\odot$) at $z = 1.299$.} Solid, multi-color lines show the solutions assuming \keff\, $= \kappa_{0,\textrm{ISM}} \frac{r}{\textrm{10\, kpc}}$ (valid in global CR flux steady-state) with `advection/streaming-like' behavior subsumed into `diffusion'. In both panels, the colors of each solid line are scaled to the corresponding \keff\, (increasing from blue to coral) at a given radius. Different colored dashed lines (blue, brown, coral) represent lower to higher constant $\kappa_{\mathrm \|}$ + \veff\, * r diffusion + advection/streaming solutions at fixed "streaming/advective" radius r$_{\textrm{st}}$ = 10 kpc.  \textbf{Left:} Solutions for a constant CR injection history with $\epsilon_{\mathrm{CR,BH}} = 3 \times 10^{-4}$.  \textbf{Right:} The same solutions for a bursty, time-dependent CR injection history for a simulated halo of the same mass with the same injection efficiency. While the steady state approximation of \keff\, $\sim r$ behavior holds out to the effective diffusive radius $\sim \sqrt{\kappa_{\textrm{eff}} \tau}$, subsuming non-steady `advection/streaming'-like behavior into \keff\, can over-/under-predict the ``true'' P$_{\textrm{CR}}$(r) at large radii due to finite travel-time effects and diffusive ``softening'', which can be very significant for r $\gtrsim$ R$_{\textrm{vir}}$ depending on non-trivial injection histories and the values of \keff\, and \veff.}
    \label{fig:varied_kappa_numerical}
\end{figure*}

In Figure \ref{fig:varied_kappa_numerical}, we compare a simple power-law scaling of \keff\, $\sim r^{-1}$ (subsuming `streaming/advection-like' behavior into diffusion), motivated by the closed form solution to the diffusion-advection/streaming dynamics in spherically-symmetric, global flux steady state to the constant diffusion-advection/streaming solution with time dependence. In these types of solutions (as discussed in \S \ref{sec:sec2}, for a given constant $\kappa_{\mathrm 0}$ and v$_{\textrm{eff}}$, for radii r $\geq$ r$_{\textrm{st}} = \kappa_{\mathrm 0}/v_{\textrm{eff}}$, \keff\, = $v_{\textrm{eff}} * r$.

Hence, in global flux steady-state, the solutions for the constant \veff\, streaming/advection only case and a \keff\, $\sim r$ `diffusive' case become degenerate beyond the `streaming-radius', out to some effective streaming/advection cut-off radius r$_{\textrm{cut, st/adv}}$ =  \veff\, $* \, \tau_{\textrm{inj}}$. Beyond this radius, the differing effects of ``streaming/advection-like" and ``diffusive" behavior may become significant. Furthermore, while this approximation holds quite well for steadily star-forming galaxies with relatively constant CR injection rates \cite{butsky_constraining_2023,hopkins_cosmic_2025,ponnada_strong_2026}, it remains unclear how well this holds for time-variable injection, and if time-dependence can break this degeneracy between transport behavior.

To focus on the investigation of steady-state assumptions, we limit the comparison here to time-steady injection vs. the most time variable (``bursty") injection scenarios. From examining Figure \ref{fig:varied_kappa_numerical}, we see that even in the continuous injection case, subsuming a constant \veff\, into purely `diffusion-like' behavior as \keff\, $\sim r$ leads to significant differences at large radii due to finite travel time effects. This owes primarily due to the extended tail of a given `diffusion-like' Green's function solution (\S\ref{sec:case1}), where despite the location of the `diffusive' peak exactly localizing to r$_{\textrm{cut, st/adv}}$, the additional diffusive behavior at the boundary moves additional CR flux outwards. This means for r $\gtrsim$ r$_{\textrm{cut, st/adv}}$, P$_{\textrm{CR}}$ is strictly over-estimated whereas for r $\lesssim$ r$_{\textrm{cut, st/adv}}$, P$_{\textrm{CR}}$ is under-estimated. Assuming steady state and subsuming `diffusion-like' behavior into a purely `streaming/advection' solution with \veff\, $\sim \kappa/r$ would result in vice-versa under-/over-estimation of P$_{\textrm{CR}}$ in this case (i.e., one would miss extended tails of CR pressure where they ought to be).

In the case examined in Fig. \ref{fig:varied_kappa_numerical}, this over-estimation is evident for the lowest \keff, \veff, parameterization in the constant injection case, and when $\tau_{\textrm{inj}}$ $>$ t$_{\textrm{adv}}$, t$_{\textrm{diff}}$ as in the higher \keff\, and \veff\, parameterizations, the steady-state \keff\, $\sim r$ begins to under-predict the true P$_{\textrm{CR}}$ at large radii. These time-dependent effects are further exacerbated for non-steady injection as evinced by the solutions for the bursty injection histories. The degree of under-/over-estimation here also depends on the time variability of CR injection (which for fixed transport parameterizations will affect the degree to which P$_{\textrm{CR}}$(r) is in steady-state), radial variation of \keff\, and/or \veff\, (which we stress can in principle arbitrarily vary with plasma conditions, but are not explored as such here), and subsequently the exact value of r$_{\textrm{st}}$, and so we simply show a characteristic case of fixed r$_{\textrm{st}}$ = 10 kpc for plausible values of constant \keff\, and \veff\, as one example. 

We again emphasize that the true behavior of CR transport at these large halo radii is essentially unknown. Direct simulations of CR pitch-angle scattering in turbulence in various regimes have rarely focused on the CGM \citep[though see][for a relevant exploration of scattering conditions in the ICM]{reichherzer_efficient_2025}. There are, however, observational hints towards rising \keff\, in the CGM of $\sim L^{\ast}$ galaxies \citep{butsky_constraining_2023}, and if \keff\, were sufficiently large and arising from the `diffusive' limit of the CR scattering rate, these `diffusive softening' effects relative to the ``true'' constant diffusion + streaming/advection solution may become less important as sharp gradients would be smoothed over and P$_{\textrm{CR}}$ at a given radius would more quickly reach a steady-state. So in this regime, the ``softened'' solution may actually present the more physically ``correct'' representation of P$_{\textrm{CR}}$ compared to the constant diffusion + advection/streaming transport parameterization. 

\begin{figure}
    \centering
    \includegraphics[width=1.0\linewidth]{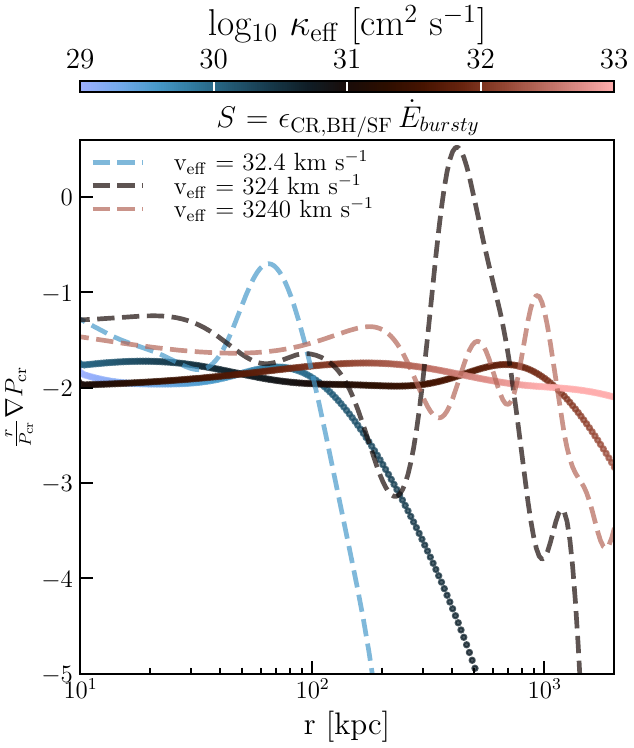}
    \caption{\textit{Dimensionless local logarithmic CR Pressure Gradient in a massive galactic halo ($M_{\textrm{halo}}^{\textrm{z=0}}$ = 10$^{13} M_\odot$) at $z = 1.25$.} Solid multi-color lines show the solutions assuming \keff\, $= \kappa_{0,\textrm{ISM}} \frac{r}{\textrm{ 10\, kpc}}$ (valid in steady-state for continuous injection) with `advection/streaming-like' behavior subsumed into `diffusion'. The colors of each solid line are scaled to the corresponding \keff\, (increasing from blue to coral) at a given radius. Different single-colored dashed lines (blue, brown, coral) represent lower to higher constant $\kappa_{\mathrm 0}$ + \veff\, * r diffusion + advection/streaming solutions at fixed streaming radius r$_{\textrm{st}}$ = 10 kpc. Use of steady-state approximations can shift strong gradients relative to the diffusion + advection/streaming solutions and smooth over features which may appear in advection/streaming-dominated transport regimes.}
    \label{fig:gradients}
\end{figure}

Nonetheless, in Figure \ref{fig:gradients}, we demonstrate how the different behaviors of the solutions in Figure \ref{fig:const_kappa_models_numerical} lead to strongly varying gradients, particularly when finite travel time effects are significant. Here, we show the dimensionless gradient of P$_{\textrm{CR}}$ at t $=$ 5 Gyr ($z=1.25$, closer in time to the peak variability of injection) for the same range of CR transport parameterizations as in Fig. \ref{fig:varied_kappa_numerical}. For fixed r$_{\textrm{st}}$, the spatial scale at which strong CR pressure gradients emerge due to finite travel time effects is strongly dependent upon how the transport is modeled i.e., using the degenerate steady-state solution vs. evolving `diffusive' and `advection/streaming-like' terms separately. Again, while we neglect adiabatic effects here, we caution that not accounting for varying source injection times will artificially modulate the physical scale at which CRs dynamically drive winds, as well as the magnitude of this effect. The significance of this time-dependent effect depends on the ratio of the $\nabla P_{\textrm{CR}}$ to the virial pressure gradient at a given radius -- here we have presented the gradients in dimensionless form as the relevance of this effect will depend on the exact energetics of the system being modeled. But we stress that in any steady-state sub-grid model, these effects may compound for non-trivial time-variable source injection and/or source spatial distribution non-linearly in the emergent dynamical effects of CRs on the background gas.

Notwithstanding a clear understanding of the CR transport, we stress that the most common assumptions in the literature thus far have invoked constant transport parameterizations as we have exemplified here, and so we caution careful consideration of these time-dependent effects in sub-resolution or analytic modeling of CR feedback around galaxies, particularly for complex source distributions or in large cosmological volumes. 
Another important consequence of our findings is that \textit{time dependence of P$_{\textrm{CR}}$ can break the degeneracy between effective `diffusion-like' and `streaming-like' behaviors,} as we aimed to discern. We discuss the observational implications of this further below.

\section{Validation of our simplified approach against full CR-MHD simulations}\label{sec:validation}

\begin{figure}
    \centering
    \includegraphics[width=\linewidth]{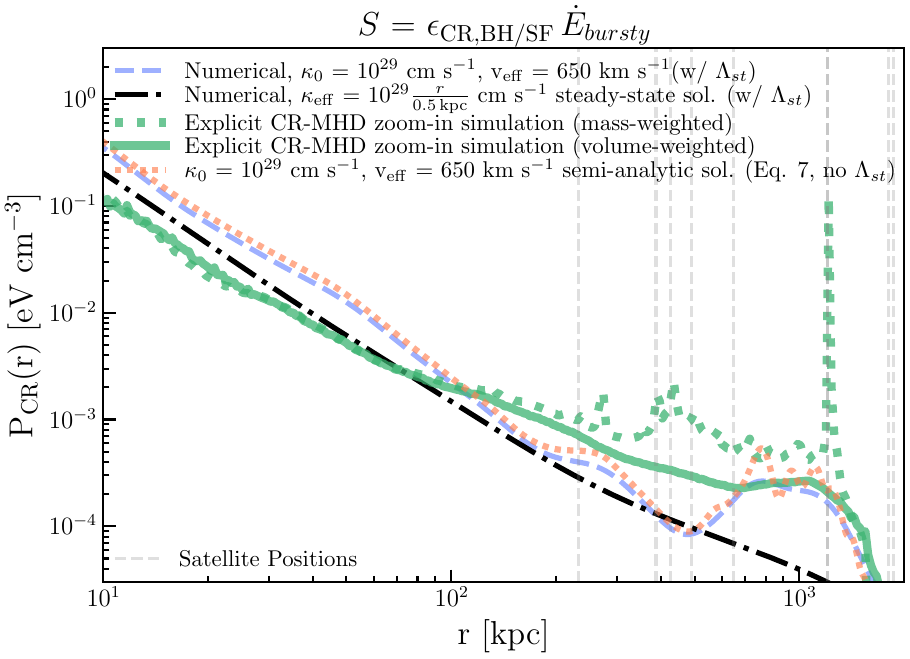}
    \caption{\textit{Validation of our simplified modeling of P$_{\textrm{CR}}$ against a fully dynamic, cosmological zoom-in simulation of a massive galactic halo ($M_{\textrm{halo}}^{\textrm{z=0}}$ = 10$^{13} M_\odot$) at $z = 1.299$.} The line for a well-fit constant \knaught\, and \veff\, numerical solution is shown in dashed blue, and the full simulation result, mass and volume weighted medians in each radial bin, are shown by the green solid and dotted lines respectively. The black dot-dashed line shows the same source injection solved numerically using the steady-state \keff\, formulation (\keff\, $\sim r$), and the orange densely dotted line shows our simplified semi-analytic approach. Accounting for time-dependence of injection, our simplified approach can reasonably match the true result at large r $\gtrsim R_{\textrm{vir}}$ to within a factor of a few, compared to the order-of-magnitude difference of the steady-state assumption at large radii. The mass-weighted simulation profile highlights the increasingly important contribution of in-spiraling satellite galaxies at large radii for modest $z$, and we indicate the positions of these with vertical grey dashed lines. These correspond with large `spikes' of CR pressure.} 
    \label{fig:numerical_validation}
\end{figure}

In Figure \ref{fig:numerical_validation}, we demonstrate the validity of our simplified assumptions and numerical modeling here by comparing our time-dependent semi-analytic and numerical solutions against a cosmological zoom-in simulation of a massive galaxy halo with explicit CR-MHD. We emphasize that it is not our goal in this work to exactly one-to-one match the reference simulation profile, as we have neglected inter alia collisional and adiabatic loss terms in our modeling for simplicity, but to demonstrate robustly capturing time-dependent effects which would otherwise be missed in a steady-state formulation. As such, we again choose a high-$z$ snapshot time of the same massive galaxy simulation as this is the regime where these effects would be most significant, and contribution from in-spiraling satellite galaxies is relatively lower.

We see that solving for P$_{\textrm{CR}}$ utilizing the star formation history, black hole accretion rate, and emergent value of \knaught\,\footnote{Here, we take \knaught\, for the $\sim$GeV proton energy bin -- as the comparison simulations are spectrally-resolved, the ``true" \knaught\, would be $\kappa_{\textrm{iso}}\, \sim$ c$^{2}$/(9 $\langle\nu_{\textrm{CR}}(r)\rangle_{E_{CR}}$), where $\langle\nu_{\textrm{CR}}(r)\rangle_{E_{\textrm{CR}}}$ is the CR energy averaged scattering rate across energy bins for gas cells at a given radius r, to account for possible shifting of the CR energy peak away from the canonical $\sim$GeV \citep{hopkins_first_2022,girichidis_spectrally_2022_paperII}. For our proof-of-concept here, we choose the strictly constant, unweighted value, as this change only marginally affects our results.} evolved directly by the simulation from the scattering rate, we are able to well reproduce the shape of the full simulation P$_{\textrm{CR}}$ profiles at large radii, which would be missed utilizing the steady-state `effective diffusion' approximation.

In Figure \ref{fig:numerical_validation}, we have assumed only streaming losses at fixed v$_{\textrm{A}}$ = 30 km s$^{-1}$ for our numerical solutions, and no losses for the semi-analytic solution following Eq. \ref{eqn:SAM_econs}. In the full simulation, all relevant loss terms are evolved. While we assume f$_{\textrm{cal}}$ = 0  for simplicity, we note that some non-zero fraction of the CR energy in the full simulation has to be lost to hadronic collisions as the central region around the BH is above the calorimetric surface density limit \citep{lacki_physics_2010,Hopkins2020,hopkins_testing_2021,ponnada_hooks_2025} at some times around  z $\sim 2-3$, but comparison of the total energy in the simulation volume to the total energy injected at the presented snapshot indicates that this is a negligible correction (i.e, a very large fraction of the CR protons indeed escape into the halo for the CR transport model in the reference simulation).


Our simplified model predictions agree quite well with the true simulation for \veff\, = 650 km s$^{-1}$, with streaming losses having negligible effect for large \veff\, (c.f. the semi-analytic line of Eq. \ref{eqn:SAM_econs} and the numerical solution lines). As we discussed before, the local Alfv\'en speed and bulk gas motions can vary arbitrarily -- nonetheless, we find that even profiles that significantly differ from the steady-state solutions can be well `fit' (to within a factor of a few) particularly at large radii by our approximations, for some constant value of \veff\, within the range we have sampled, particularly at large radii r $\gtrsim R_{\textrm{vir}}$ where the bulk of the CR energy resides in this example case, and indeed capture features owing to extended periods of enhanced injection at higher $z$ transported to the outer halo via ``streaming/advection-like" behavior. Such features around r $\sim$ 1000 kpc ($\sim$ 3 R$_{\textrm{vir}}$) are notably missed by the steady-state \keff\, formulation, and as we stress in the earlier section, these can be quite significant for the dynamical effects of CRs on the background gas and subsequent evolution of the galaxy. Though, there are still some noteworthy differences particularly at the inner radii, which might arise from not modeling adiabatic losses or from the approximation made in Eq. \ref{eqn:SAM_econs} for the numerical and semi-analytic solutions, respectively.

We note also the mass weighted profile of the ``true" simulation shows significant `bursts,' several of which are coincident with the positions of in-spiraling satellite galaxies at that snapshot. While we compare against the volume-weighted median to avoid biasing our validation (particularly as we do not include those satellite galaxies as sources in our modeling), we emphasize that the CR contributions from these satellites as they similarly diffuse and stream/advect outwards may contribute to the volume-weighted lines, hence explaining why we predict coincident `dips' in our model lines evolving solely the central galaxy contributions. We stress that at lower redshifts, as these satellite galaxies of massive halos will have rising or approximately constant star formation histories relative to the massive central, which is expected in a population-averaged sense \citep{zhang_trinity_2023}, these contributions may become increasingly important for massive groups and clusters at large halo radii from the central massive galaxy \citep[see also][]{roy_survive_2025}.

\section{Discussion and Conclusions}\label{sec:discussion_conclusions}

In this work, we have analytically and semi-analytically detailed the effects of arbitrarily time-variable CR injection and spatially constant transport parameters on the bulk CR pressure/energy distribution in galaxy halos. We demonstrate that bursty and time-dependent sources can have substantial effects on the distribution and evolution of CR pressure, particularly on large scales which may not be in pressure steady-state, and for which finite travel-time effects and/or source evolution can modify solutions significantly from steady-state expectations. Below, we summarize our results and discuss their relevant implications.

\subsection{Implications for sub-grid models of CR feedback and outlining a new approach}\label{subsec:subgrid_considerations}

We stress that these time-dependent effects are particularly of concern when utilizing sub-grid models for CR feedback in large volume simulations. For instance, if we were to utilize the steady-state formalism of \citet{hopkins_simple_2023}, Figures \ref{fig:varied_kappa_numerical} and \ref{fig:gradients} make clear how at fixed simulation time, one may over-predict the true P$_{\textrm{CR}}$ at a given radius if not accounting for finite travel-time effects of a given source (by implicitly using a larger than physical r$_{\textrm{max}}$ or equivalently t$_{\textrm{max}}$ in the sub-grid model proposed there), while simultaneously missing or shifting strong gradients owing to this same truncation due to diffusive `softening' effects if the transport is in the `streaming/advection-like' regime. 

In this paper we have only exemplified how these behaviors manifest at fixed time intervals, but we emphasize that these effects would compound non-linearly in any sub-grid implementation of CR feedback coupled to hydrodynamics, and it is not trivial to understand exactly how. Even in the example above, if one were to over-predict P$_{\textrm{CR}}$ but miss large $\nabla P_{\textrm{CR}}$, it is unclear whether one would still find the ``true" dynamical outflow/inflow condition for the same conditions at a given time (thermal/magnetic pressure at a given radius etc.) but for the wrong reasons, or simply find the wrong answer entirely. This is particularly of importance as the non-linear, dynamical interaction of CRs in galactic halos via large-scale pressure gradients is the \textit{principal} method by which several studies in the literature find CRs impact galaxy formation \citep{butsky_role_2018,Hopkins2020,buck_effects_2020,hopkins_effects_2021,huang_launching_2022,quataert_physics_2022,armillotta_cosmic-ray_2024}. Missing these effects across large time intervals would thus compound non-linearly with other generic galaxy formation physics of gas dynamics, star formation, and stellar/black-hole feedback via thermal and mechanical channels.

So, we recommend that `best practice' for sub-grid implementations carefully consider the finite time-travel effects of each independent source, evaluating the contributions from each source at $\tau_{\textrm{inj},i}$, the time since injection from each source. As \citet{hopkins_simple_2023} note, this presents inherent challenges for implementation in tree-based codes, and so we leave development of fast numerical schemes for updates to such sub-grid models to future work. The proof-of-concept of the semi-analytic, time-dependent approach we demonstrate here presents a new avenue to create a new subgrid model for CR feedback.

While we leave presentation of a comprehensive numerical recipe and validation across different galaxy mass regimes to future work, we briefly outline the approach here. For each source, P$_\textrm{CR}$($\vec{r}$,t) can be estimated via Eq. \ref{eqn:PdotCR_const_diff_adv_approx} at a given timestep and appropriately normalized following the time-dependent kernel in Section \ref{sec:case3}, for a given choice of \knaught\, and/or \veff \,(with \veff\, potentially informed by the mean outflow velocity field in the simulation volume). To minimize the overhead cost of tracking each individual source, one could group sources (star forming particles or black hole sink particles) within each node to avoid tree-recomputation and compute the contribution to P$_\textrm{CR}$($\vec{r}$,t) from each node. 

For each gas cell, the relevant tree nodes can be flagged to avoid unnecessary computation. For example, in large volumes containing massive galaxies or clusters, the contribution to P$_\textrm{CR}$ at very large radii ($\gtrsim$ R$_{\textrm{vir}}$) can become dominated by satellite galaxies at late times owing to quenching of centrals in contrast to rising SFHs in satellites. Every few time-steps (which requires calibration), the source tagging can be re-identified to avoid artificially missing source contributions owing to the prior grouping. Then, the relevant ``shielding" terms to account for losses can be computed post-hoc, akin to the LEBRON methods outlined in other subgrid models \citep{hopkins_simple_2023} and the relevant CR thermochemical, pressure, and radiative contributions can be conjoined with the hydrodynamics, at which point the adiabatic exchange can also be estimated.

\subsection{Implications for determining `streaming/advection-like' vs. `diffusive' transport in galaxy halos}

In \S \ref{sec:sec3}, we discussed how for constant CR transport parameters, the `diffusion-like' and `streaming-like' solutions become degenerate in pressure steady-state given a time-steady source term. We find that for non-trivial source injection, deviation from the steady-state formulation can break this degeneracy. As we have noted above, in this work we are agnostic to whether the behavior in galaxy halos, particularly in the outer halos where there are few constraints, is `streaming-like' vs. `diffusion-like'. Given that we have shown the predictions diverge owing to time-dependent effects, this opens a potentially interesting avenue to constrain the effective transport behavior in halos.

In Figure \ref{fig:numerical_validation}, we show how the steady-state formulation of subsuming `advective/streaming-like' behavior into `diffusion' with \keff\, $\sim r$ can lead to flattening out features associated with elevated injection over finite time intervals at higher redshifts (as is expected for virtually any massive galaxy with strongly peaked BH accretion/star formation at $z \sim 2-3$) which otherwise are only modestly smoothed during advective/streaming-like transport out to large radii. 

We speculate that such transport features may be related to observable phenomena around massive galaxies and clusters, and are potentially of significance for constraining CR transport. For instance, `Odd Radio Circles,' or ORCs, \citep{Norris_2021_ORCs,Norris_2021_review} which are edge-brightened discs of diffuse radio continuum emission have been detected in recent $\sim$ GHz surveys. Follow-up observations have spatially associated these objects on-sky with massive galaxies $ M_{\textrm{halo}}\sim 10^{13} M_\odot$  at $z \sim 0.2-0.6$. While some studies have proposed these to be associated with virial shocks around such galaxies, \citep{yamasaki_are_2023}, there have also been idealized simulations demonstrating that CR-laden AGN jets can reproduce similar morphological features \citep{lin_agn_2024} -- indeed heuristically the features we see naturally emerge from bursty SF/AGN injection transported out via streaming/advection in Figure \ref{fig:numerical_validation} bear resemblance to such structures at large radii, albeit at different redshift. 

We leave an extensive survey of injection + transport conditions to future work, but we note that for plausible injection efficiencies, accretion/SF histories, and bulk CGM CR transport speeds, such features may represent transient periods in the bulk propagation of CRs out to large radii, which may inform us about the nature of CR transport in the outer CGM of massive galaxies. Therefore, the observed rarity of ORCs may be related to balance of diffusive vs. streaming-like transport with losses (and possible re-acceleration) out to a given galactocentric radius convolved with a time-dependent source injection history. For instance, in the event of a strong but transient starburst-driven wind (as invoked to explain the ORC modeled in \cite{coil_ionized_2023}) the ratio between the `diffusive' and `advective/streaming'-like speeds in massive galaxy halos heuristically may determine the size and longevity of an edge-brightened, ring like structure (c.f. the survival or absence of over-pressurized regions in Figs. \ref{fig:varied_kappa_numerical} and \ref{fig:gradients}).

\subsection{Motivation for exploring CR feedback across galaxy mass scales}

In the 2nd row of Figure \ref{fig:const_kappa_models}, we demonstrated how even modest `advection/streaming-like' behavior and time-dependent injection shifts P$_{\textrm{CR}}$ out to larger radii and therefore significantly shifts the distribution of E$_{\textrm{CR}} \sim r^2 P_{\textrm{CR}}$ outward relative to steady-state approximations. This further corroborates the findings of \citet{quataert_cosmic_2025}, who demonstrate for single $\delta$-injection at $z \sim 2-3$ and plausible \knaught\,, $P_{\textrm{CR}}$ can be dynamically significant in the outskirts of galaxy groups and clusters and potentially resolve the $\sigma_{8}$ tension between weak-lensing and cluster SZ measurements \citep{des_collaboration_dark_2022}. We emphasize that not only may this be the case, but time-dependent effects may considerably compound, subsequently affecting how CR feedback influences both the matter power spectrum and regulation of galaxy growth via `preventative' feedback on large-scales ($k^{-1} \sim \textrm{Mpc}$). 

Idealized simulations of CR jets \citep{su_which_2021,su_unraveling_2023,lin_agn_2024,su_modeling_2025} and cosmological zoom-in simulations with BH-generated CRs and explicit CR-MHD \citep{wellons_exploring_2023,byrne_effects_2024,ponnada_hooks_2025} are limited in their predictive power at these large-scales, therefore implementation of CR feedback in large volume cosmological simulations of galaxy formation \`a la IllustrisTNG \citep{pillepich_simulating_2018},  Simba \citep{dave_simba_2019}, Astrid \citep{ni_astrid_2022}, FABLE \citep{henden_fable_2018}, and FIREbox \citep{feldmann_firebox_2023} etc. is necessary. 

Towards this end, \citet{ramesh_illustristng_2024} implement the steady-state sub-grid formulation of \citet{hopkins_simple_2023} in a (25 Mpc h$^{-1}$)$^{3}$ cosmological volume utilizing TNG physics, incorporating CR contributions from stellar sources. The implementation therein utilizes t$_{\textrm{max}}$ = t$_{\textrm{sim}}$, which as we caution in \S \ref{subsec:subgrid_considerations} would over-estimate the true pressure contribution from all sources of finite age due to finite travel-time effects and potentially miss or shift strong  pressure gradients, in addition to the other limitations described in \citet{hopkins_simple_2023} regarding neglecting adiabatic losses. Notwithstanding these caveats, \citeauthor{ramesh_illustristng_2024} find qualitatively similar conclusions regarding CR effects to the literature results from zoom-in and idealized simulations of $\lesssim L^{\ast}$ galaxies, warranting further study. 

We propose further exploration of CR effects in large volumes akin to \citet{ramesh_illustristng_2024} (particularly even larger volumes which would more statistically sample galaxy groups and clusters) including  treatments of CR injection from AGN (e.g., including small fixed-fractions of accretion energy in the form of a CR fluid as per \citep{wellons_exploring_2023}) as this would potentially expand the prediction space for CR effects via additional observables probing more massive galaxy groups and clusters, e.g. weak-lensing and other matter clustering constraints \citep{sharma_hydrodynamical_2025}, X-ray emission and tSZ stacks \citep{zhang_hot_2024,das_thermal_2025,ponnada_strong_2026}, and radio continuum observations \citep{van_weeren_diffuse_2019}. 

And of course, comparisons of such observations with crudely parameterized theoretical predictions will only net an understanding of `effective' bulk CR transport parameters averaged over large spatial scales, but these present a way to empirically constrain the resurging theoretical study of CR scattering in MHD turbulence from first-principles \citep{kempski_cosmic_2023,lemoine_particle_2023,kempski_unified_2024,butsky_galactic_2024} and thus the emergent effects on galaxy formation across cosmic epochs and mass scales, towards a clear physical theory of CR transport in galactic environments.

\section*{Acknowledgements}
We wish to recognize and acknowledge the past and present Gabrielino-Tongva people and their unceded Indigenous lands upon which this research was conducted. SP thanks the two anonymous referees for their constructive comments that significantly improved this manuscript. SP thanks Phil Hopkins, Iryna Butsky, and Charvi Goyal for useful discussions leading to the generation of this manuscript. Support for SP was provided by NSF Research Grants 20009234, 2108318, NASA grant 80NSSC18K0562, and a Simons Investigator Award. Numerical calculations were run on NSF/TACC allocation AST21010, TG-AST140023, and TG-PHY240164.

%

\vspace{5mm}





\bibliography{Time_dependent_CRs}{}
\bibliographystyle{apsrev4-2}



\end{document}